\documentclass[]{aa}
\usepackage{amsmath}
\usepackage{graphicx}
\begin{document}
\headnote{Research Note}
\title{Breaking the core-envelope symmetry in p-mode pulsating stars}
\titlerunning{Breaking the core-envelope symmetry}
\author{A. Miglio\inst{1,2}, E. Antonello\inst{2}}
\institute{Institut d'Astrophysique et de G\'eophysique de l'Universit\'e de Li\`ege, All\`ee du 6 Ao\^ut, 17 B-4000 Li\`ege, Belgium \and INAF-Osservatorio Astronomico di Brera, Via Bianchi 46, 23807 Merate, Italy }

\offprints{A. Miglio,\\ \email{miglio@astro.ulg.ac.be}}

\abstract{It has been shown that there is a potential ambiguity in the asteroseismic determination of the location of internal structures in a pulsating star (\cite{montgomery}). We show how, in the case of high-order non-radial acoustic modes, it is possible to remove this ambiguity by considering modes of different degree.
To support our conclusions we have investigated the seismic signatures of sharp density variations in the structure of quasi-homogeneous models.
}
\maketitle

\section{Introduction}
Asteroseismology is finally providing strong and independent constraints to the modeling of several classes of pulsators, such as $\beta\;  Cephei$ stars (\cite{aerts}), solar-like stars (\cite{thoul}, \cite{dimauro}) and white dwarfs (e.g. \cite{metcalfe}).
The forthcoming observations of stellar pulsations from space (MOST and COROT) will detect low-degree acoustic modes with sufficient accuracy to allow the determination of local characteristics of stellar interiors.

It is well known that a sharp variation in the equilibrium structure of a star gives rise to a periodic component in the frequencies of oscillation (see for example \cite{monteiro}). A way to isolate these components, in high order modes, is to consider deviations from asymptotic expressions for period (frequency) spacing in g-mode (p-mode) pulsators.

It has been recently shown by Montgomery et al. (2003) that in the case of white dwarfs, where only high order gravity modes have been detected, there is a potential ambiguity in determining where in the stellar interior  the variation that generates the periodic signal is located. With the aim of extending the analysis to acoustic modes, we present how it is possible to remove such an ambiguity by considering modes of different degree.

In Section \ref{sec:alias} we show the asymmetry between signals generated in the core and in the envelope.  The seismic signatures of sharp density variations in quasi-homogeneous models are presented in \ref{sec:models}. Lastly we give our conclusions in Section \ref{sec:conclusions}.

\section{Aliasing}
\label{sec:alias}
We can estimate the contribution to the p-mode oscillation spectrum of localized density variations (a discontinuity in the first derivative of density) following the approach by \cite{monteiro94} (MCDT in the following).

This approach could be generalized to account for a `bump' in a convenient equilibrium variable describing a sharp feature in the stellar interior, e.g. the first adiabatic exponent $\Gamma_1$ if analyzing the second helium ionization zone (\cite{monteiro98}), or derivatives of the sound speed if studying the characteristics of the base of convective envelopes in solar-type stars (\cite{monteiro}). 

The contribution to the oscillation spectrum of such a sharp variation could be estimated as the periodic component of the difference ($\delta\nu$) between the acoustic frequencies of the star showing such a sharp variation and the frequencies of an otherwise fictitious smooth model.

It is possible to express $\delta\nu$ as
\begin{equation}
\delta\nu=\dfrac{\delta I}{I_1}
\label{eq:perio}
\end{equation} 
where $I_1$ depends on the structure of the smooth model at equilibrium and on its eigenfunctions, and $\delta I$ depends also on the characteristics of the sharp variation (see MCDT for details).

We expressed $\delta I$ only in terms of the perturbation $\delta\rho$ using
\begin{equation}
\dfrac{\delta \rho}{\rho}=\dfrac{\delta(\Gamma_1 p)}{\Gamma_1 p}-\dfrac{\delta c^2}{c^2}
\end{equation}
and neglected then terms in $\delta(\Gamma_1)$ as we consider $\Gamma_1$ uniform throughout the model. The contribution from terms in  $\delta(p)$ could also be neglected since a discontinuity in the $n$-th derivative of $\rho$ will generate a discontinuity only in the $n+1$-th derivative of the pressure $p$, and, as shown below, the dominant term in $\delta\nu$ comes from the lowest order discontinuous derivative of an equilibrium variable.

For high order modes we can therefore write
\begin{equation}
\delta\nu_p\simeq\int_{0}^{T}{\left[\dfrac{a_1(\tau)}{\nu^2}\dfrac{\rm d^3 }{\rm d\tau^3}\left(\dfrac{\delta \rho}{\rho} \right)+\dfrac{a_2(\tau)}{\nu^2} \dfrac{\rm d^2 }{\rm d\tau^2}\left(\dfrac{\delta \rho}{\rho} \right)\right]\sin{(\Delta)}\rm d \tau},
\label{eq:var}
\end{equation}
where for low-degree modes and a discontinuity located far from the turning points of the mode, $\Delta\simeq2\pi\,\nu\,2\tau+2\phi$, where $\tau$ is defined as
$$
\tau(r)=\int_r^R{\frac{{\rm d}r'}{c(r')}}, 
$$
$T=\tau(0)$ and where $a_1(\tau)$ and $a_2(\tau)$ depend only on the structure of the smooth model.

We have considered the density discontinuous in its first derivative, thus we need to integrate by parts the term in the third derivative of density and, as a first approximation, keeping track only of terms of order $O(\nu^{-1})$, we find
\begin{equation}
\delta\nu_p\simeq \dfrac{a}{\nu}\sin{(2\pi\,\nu\,2\tau_d+\phi')},
\label{eq:signal}
\end{equation}
where $a$ is evaluated at the acoustic depth $\tau_d$ of the discontinuity considered.

When looking for such a periodic signal in the frequencies of an acoustic oscillation spectrum it is clear that the signal could be evaluated only in a discrete set of frequencies $\nu_{\rm n,\ell}$. 

Let us consider modes of same degree $\ell$ and a periodic signal as in Eq.(\ref{eq:signal}).
Having defined the acoustic radius as 
$$
\theta(r)=\int_{0}^{r}{\frac{{\rm d}r'}{c(r')}}
$$
and remembering the simple asymptotic relation (\cite{tassoul}) 
\begin{equation}
\nu_{n}\simeq\Delta\nu\,\left(n+\phi' \right) 
\label{eq:disp}
\end{equation} 
and
\begin{equation}
T\equiv\tau(0)\equiv\theta(R)\simeq 1/(2\Delta\nu)
\end{equation} 
it is straightforward to show that $\delta\nu$ could also be written as
\begin{equation}
\delta\nu=A(\nu)\sin{(2\pi\,\nu\,2\theta_d+\phi'')}
\end{equation}
This means that we cannot distinguish whether a variation is located at an acoustic depth $\tau$ or $T-\tau$ (see Mazumdar \& Antia, 2001; see Montgomery et al., 2003,  in the case of g modes).
\begin{figure}
\resizebox{\hsize}{!}{\includegraphics[angle=90]{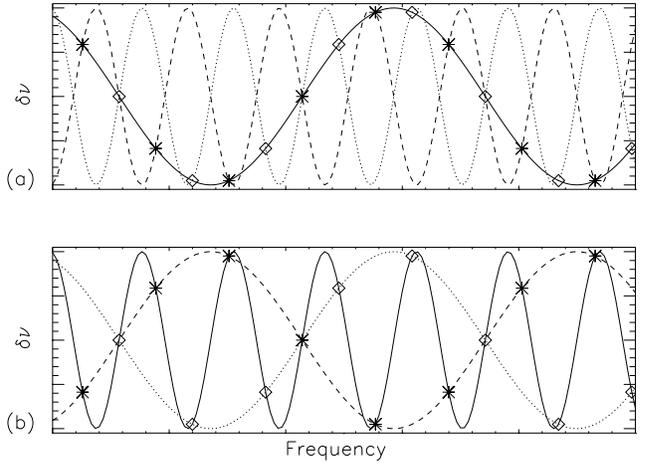}}
\caption{(a) The discontinuity is located near the surface. The periodic signal could be described as slowly changing and independent from $\ell$, with a period $\sim 1/(2\tau)$ (continuous line) or as a signal with a short period $(1/(2\theta))$ with a phase which depends on $\ell$ being even or odd (dashed and dotted lines). The signal evaluated at the discrete frequencies in Eq. (\ref{eq:tassoul}) is represented by asterisks ($\ell = 0$) and diamonds ($\ell = 1$). The values on the axes are arbitrary. (b) The discontinuity is located near the center of the star.}\label{fig:alias}
\end{figure}

Since we would like to include in our treatment modes of different degree $\ell$ we generalized this argument considering the more general expression (Tassoul, 1980)
\begin{equation}
\nu_{n,l}\simeq\Delta\nu\,\left(n+\frac{\ell}{2}+\phi' \right) 
\label{eq:tassoul}
\end{equation} 
instead of Eq. (\ref{eq:disp}). It follows therefore that the signal in Eq. (\ref{eq:signal}) is equivalent to 
\begin{equation}
\delta\nu=(-1)^\ell A(\nu)\sin{(2\pi\,\nu\,2\theta+\phi'')}
\label{eq:newperio}
\end{equation} 
We can conclude that, to this first approximation, it is equivalent to consider a periodic signal independent from $\ell$ with a ``frequency'' twice the acoustic depth of the discontinuity or signals that have a ``frequency'' twice the acoustic radius of the discontinuity and depend on $\ell$ through the multiplying factor $(-1)^\ell$. This can be applied both to periodic signals generated by discontinuities located near the surface and the center of the star (see Fig. \ref{fig:alias}).

\section{Signatures of density variations in quasi-homogeneous models}
\label{sec:models}
As a first investigation we considered models with simple density profiles, we derived analytically the pressure profile and calculated acoustic oscillation spectra using \emph{Aarhus Adiabatic Pulsation Package}\footnote{\texttt{http://astro.phys.au.dk/\~{}jcd/adipack.n/}}. 
We looked for oscillatory signals in the large frequency separation $\Delta\nu$ where 
$$\Delta\nu_{n\ell}=\nu_{n+1,\ell}-\nu_{n,\ell} $$
and in $D_{n\ell}$, a measurement of the small separation defined as
$$D_{n\ell}=(\nu_{n,\ell}-\nu_{n-1,\ell+2})/(4\ell+6)$$

We considered sharp variations in $\rho$ located at two symmetric normalized acoustic depths $\tau_{\rm a}/T$ and $\tau_{\rm b}/T$ (i.e. $\tau_{\rm b}/T=1-\tau_{\rm a}/T$ ).

\begin{figure}
\resizebox{\hsize}{!}{\includegraphics[angle=90]{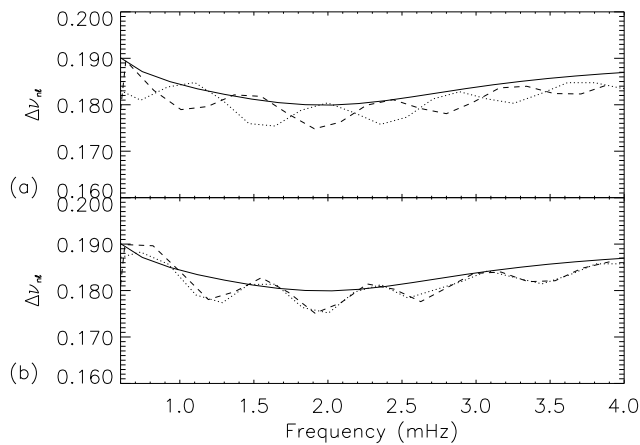}}
\caption{Large separation as a function of frequency. The continuous line is  $\Delta\nu_{n0}$ ($=\Delta\nu_{n1}$) for the uniform mass model; the dashed and dotted lines represent respectively $\Delta\nu_{n0}$ and $\Delta\nu_{n1}$ for models with a sharp density variation at (a) $\tau_a/T=0.8$ and (b) $\tau_b/R=0.2$.}\label{fig:large}
\end{figure}
\begin{figure}
\resizebox{\hsize}{!}{\includegraphics[angle=90]{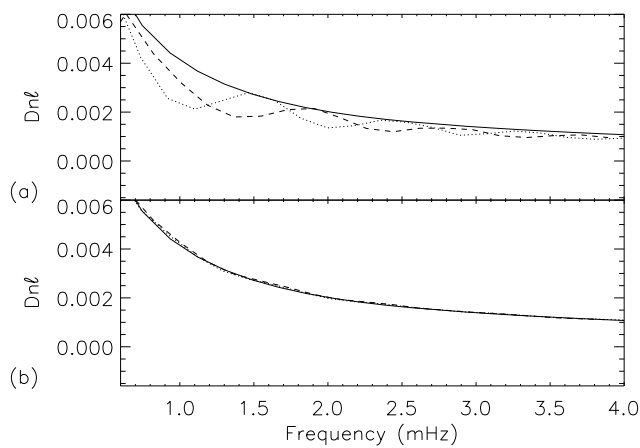}}
\caption{Small separation as a function of frequency. The continuous line is  $D_{n0}$ (=$D_{n1}$) for the uniform mass model; the dashed and dotted lines represent respectively $D_{n0}$ and $D_{n1}$ for models with a sharp density variation at (a) $\tau_a/T=0.8$ and (b) $\tau_b/T=0.2$.}\label{fig:small}
\end{figure}

As presented in Fig. \ref{fig:large}, if we consider a model with uniform mass density, $\Delta\nu_{n\ell}$ does not depend on the degree $\ell$ and has a smooth behaviour throughout the range of frequencies considered. If we introduce a density variation in the central regions of the equilibrium model (a step function in the derivative of the density profile at an acoustic depth $\tau_{\rm a}$) we notice the appearance in $\Delta\nu_{n\ell}$ of a periodic signal  whose phase depends on the degree, as qualitatively predicted by Eq. (\ref{eq:newperio}). If the sharp variation is displaced to an acoustic depth $\tau_{\rm b}=T-\tau_{\rm a}$ (i.e. near the surface of the star) the period of the oscillatory component is aliased to the one of the signal generated by a sharp feature in the core but no dependence on $\ell$ is present.

From the analysis of $D_{n\ell}$ presented in Fig. \ref{fig:small} it becomes clear that this seismic indicator, as predicted by the asymptotic theory (\cite{tassoul}), is mostly sensitive to the central regions of the star: whereas a periodic signal appears in $D_{n\ell}$ when a discontinuity near the core is considered, no significant periodic signal is detectable if the sharp variation is displaced to its symmetric point near the surface.

The simple analysis presented here makes clear that, when considering p-mode pulsating stars, it is sufficient to consider modes of different degree or different seismic indicators to distinguish between a signal generated by a sharp variation located near the core or the surface of the star. 

As outlined in Montgomery et al. (2003) there may exist additional ways to break the core-envelope symmetry such as modeling in detail the discontinuity and obtaining an explicit expression for the amplitude of the periodic component in the frequencies of oscillation.

\section{Discussion and conclusions}
\label{sec:conclusions}
Differently than in the case of high-overtone g-modes in white dwarfs, in p-mode pulsating stars the core-envelope symmetry could be easily broken by considering modes of different degree. The ``short period'' periodic signal, signature of a sharp variation in the core a star, is expected to be aliased to a ``long period'' signal whose amplitude, in the approximation presented, depends on $\ell$ through the multiplying factor $( -1)^\ell$. This could be relevant, for example, while searching for signatures of convective cores in stars showing solar-like oscillations.


\end{document}